\documentstyle[eqsecnum,preprint,aps]{revtex}

\def\setdsize{
  \oddsidemargin=-8mm
  \evensidemargin=-8mm
  \textwidth=174mm
  \tolerance=10000
}
\def\documentformat{
  \baselineskip=20.5pt plus 1pt minus 1pt
  \parskip=5pt
}
\def\absformat{
  \small
  \baselineskip=16pt
  \noindent
}

\def\Eqn#1{(\ref{#1})}
\def\Ref#1{\!\!\!\cite{#1}}
\def\refmk#1{${}^{#1}$}
\def\figcapt#1#2{
    \caption{\protect\normalsize \label{#1}#2}
}

\def\d{{\rm d}}
\def\Id{{\bf 1}}
\def\Im{{\rm Im}}
\def\Tr{{\rm Tr}}
\def\Sup(#1,#2){#1^{\hbox{\tiny #2}}}
\def\iprod{\!\cdot\!}
\def\bra#1{\left\langle#1\right\vert}
\def\ket#1{\left\vert#1\right\rangle}

\def\osh{\omega_{\rm sh}}
\def\Nsh{N_{\rm sh}}
\def\loct{\lambda_{30}}
\def\gosc{g_{\rm osc}}

\def\ie{{\it i.e.}}
\def\eg{{\it e.g.}}

\setdsize
\begin{document}
\documentformat

\rightline{\hbox to28mm{KUNS 1229\hfill}}
\rightline{\hbox to28mm{November 1993\hfill}}
\vspace{12mm}

\begin{center}
{\large\bf
Semiclassical Analysis of the Supershell Effect\\
in Reflection-Asymmetric Superdeformed Oscillator
}
\end{center}

\begin{center}
Ken-ichiro {\sc Arita}
and Kenichi {\sc Matsuyanagi}
\end{center}

\begin{center}
{\it
Department of Physics, Kyoto University, Kyoto 606-01
}
\end{center}
\vspace{8mm}

\centerline{\bf Abstract}
\bigskip

\centerline{
\parbox{150mm}{
\absformat
An oscillatory pattern in the smoothed quantum spectrum, which is
unique for single-particle motions in a reflection-asymmetric
superdeformed oscillator potential, is investigated by means of the
semiclassical theory of shell structure.  Clear correspondence between
the oscillating components of the smoothed level density and the
classical periodic orbits is found.  It is shown that an interference
effect between two families of the short periodic orbits, called
supershell effect, develops with increasing reflection-asymmetric
deformations.  Possible origins of this enhancement phenomena as well
as quantum signatures of period-multipling bifurcations are discussed
in connection with stabilities of the classical periodic orbits.
}}

\section{Introduction}

  Possible occurrence of instability of superdeformed (SD) nuclei
having the prolate shape with the axis ratio approximately 2:1 against
the octupole-type reflection asymmetric deformation is one of the
current topics of growing interest in high-spin nuclear structure
physics.  Regions in the ($N$,$Z$) plane where we can expect existence
of reflection-asymmetric SD nuclei have been
investigated\refmk{\Ref{Dudek}\sim\Ref{Skalski}} mainly by means of
the Strutinsky-type calculations of the collective potential energy
surface (see also Refs.~$\Ref{Bonche}$ and $\Ref{SkalHe}$ for other
approaches).  Concerning the physical condition for the occurrence of
the octupole instability, Nazarewicz and
Dobaczewski\refmk{\Ref{NazDob}} have recently discussed the connection
between the dynamical symmetry of the anisotropic harmonic-oscillator
with frequencies in rational ratio and the multi-cluster
configurations.  They have suggested that the closed-shell
configurations in the prolate SD oscillator potential, defined as
having the frequency ratio $\omega_\perp/\omega_z=2$, might be
unstable (stable) against the octupole-type reflection asymmetric
shapes when the single-particle levels are filled up to the major
shells with $\Nsh$=even (odd), $\Nsh$ being the shell quantum number
defined by $\Nsh=2n_\perp+n_z$ (see also the previous work,
Ref.~$\Ref{Bengtsson}$).  Their suggestion is in good qualitative
agreement with the realistic shell-structure energy calculation by
H{\"o}ller and {\AA}berg.\refmk{\Ref{HolAb}} We call the
$\Nsh$-dependence of the octupole instability ``odd-even effect in
$\Nsh$''.

  We have suggested in Refs.~$\Ref{AriMat}$, $\Ref{MizNak}$ a possible
relationship between the odd-even effect mentioned above and the
``supershell effect'' in reflection-asymmetric SD potentials.  The
general concept of supershell was originally introduced by Balian and
Bloch\refmk{\Ref{Balian}} in relation to the semi-classical theory of
shell structure.  Quite recently, the supershell effect has been
observed, for the first time, in the mass abundance spectra of metal
clusters.  Theoretical analysis of this phenomena has been made by
Nishioka, Hansen and Mottelson.\refmk{\Ref{NisHanMot},\Ref{Nishioka}}

  As is well known, clustering of eigenvalues, that is, oscillating
pattern in the energy-smoothed level density for single-particle
motions in the mean field is called shell
structure.\refmk{\Ref{Brack}} In the semiclassical theory, classical
periodic orbits having relatively short periods are responsible for
the clustering of the levels; the frequencies in the level density
oscillation is determined by the corresponding periods of classical
motion.\refmk{\Ref{BohMot}\sim\Ref{Ozorio}} When two families of short
period orbits interfere and produce an undulating pattern in the
oscillating level density, this pattern is called supershell
structure.\refmk{\Ref{Balian},\Ref{BohMot}} In the case of the metal
clusters, a beautiful beating pattern enveloping individual shell
oscillations which is caused by the interference between the
triangular and square orbits of an electron in a spherical Woods-Saxon
potential has been demonstrated\refmk{\Ref{NisHanMot},\Ref{Nishioka}}
to nicely correspond to the experimental data.

  In the case of the SD nuclei under consideration, an interference
effect is expected to arise between the classical periodic orbits with
period $T\approx2\pi/\omega_\perp$ and those with
$T\approx2\pi/\omega_z$ of a nucleon in the reflection-asymmetric SD
potential.\refmk{\Ref{MizNak}} The main purpose of this paper is to
show that the interference effect brings about another example of the
supershell structure, which is intimately connected with the odd-even
effect in $\Nsh$ mentioned above, and which is relevant to
experimental investigations.  It should be emphasized here that,
contrary to the case of spherical potentials, the Hamiltonian
describing the single-particle motions in a deformed mean field is
non-integrable in general.  Accordingly, our Hamiltonian system is a
kind of mixed system whose phase space is composed both of regular and
chaotic regions.  As a consequence, properties of our phase space
change in a quite sensitive manner when the shape of the nuclear
surface is varied.  In this paper, we shall show that the supershell
structure becomes more significant with increasing octupole
deformation.  Possible origins of this enhancement will be discussed
in relation to the change in the properties of the classical periodic
orbits as a function of the octupole deformation parameter.

  After briefly reviewing the semiclassical theory of shell structure
in \S2, we first apply in \S3 both the torus quantization method and
the periodic-orbit quantization method to the case of the prolate SD
oscillator potential.  In this integrable limit, the supershell effect
can be treated analytically.  In \S4, a reflection-asymmetric SD
potential model is introduced and the supershell pattern in the
quantum level spectrum calculated for this potential is exhibited.  In
\S5, we investigate properties of classical motions in this potential,
like stabilities and bifurcation phenomena of the periodic orbits.  In
\S6, we show that a nice correspondence holds between peak positions
of the Fourier transform of quantum spectrum and periods of classical
closed orbits; relative heights between peaks change as functions of
the octupole-deformation parameter, providing us with a semiclassical
interpretation of the origin and the octupole-deformation dependence
of the supershell structure.  Here, quantum signature of the
bifurcations is also discussed.  Summary of this work is given in \S7.

  A preliminary version of this work was previously reported in this
journal.\refmk{\Ref{Arita}}

\section{Some Elements of Semiclassical Theory of Shell Structure}

  In this section we briefly review some basic elements of the
semiclassical theory of shell structure, which are necessary for
later discussions.

\subsection{Torus Quantization}

  To begin with, let us consider the case of multi-dimensional,
integrable Hamiltonian system, where the Hamiltonian can be written as
a function of only action variables $I_i$, being independent of angle
variables $\theta_i$ conjugate to them.  Semiclassical quantization
condition valid for such systems has been formulated by
Einstein-Brillouin-Keller, and called torus quantization or the EBK
quantization;
\begin{equation}
I_i(E_{n_1,\cdots,n_f})=\oint_i\bbox{p}\d\bbox{q}
=\hbar(n_i+\alpha_i/4)~,\hskip6mm
i=1,\cdots,f
\label{ebkcond}
\end{equation}
where indices $i$ represent mutually independent paths on
$f$-dimensional torus constructed by classical trajectories,
$\alpha_i$ are Maslov indices related to the singularities of the Van
Vleck determinant appearing in the semiclassical propagator along the
path $i$.  Thus, the semiclassical level density is given by
\begin{equation}
g(E)=\sum_{\{n\}}
\delta\Big(E-H\big(I_i=\hbar(n_i+\alpha_i/4)\big)\Big)~.
\end{equation}
The summation on the r.h.s. may be rewritten using the Poisson sum
formula into the form of topological sum over periodic
orbits.\refmk{\Ref{BerTab}} In Ref.~$\Ref{BerTab}$, spherical systems
are analyzed and clear correspondence between the topological sum and
the periodic orbits is shown.  In the spherical case, all periodic
orbits satisfy the resonance condition, \ie, the frequency ratio of
radial and angular motions are the same as that of topological
indices.  We shall apply this method to the SD harmonic oscillator
potential in \S3, and discuss the correspondence between the
topological indices and periodic orbits.  There, it will be shown that
some ``partially-resonant'' terms play an important role giving rise
to the supershell effect.

\subsection{Periodic-Orbit Quantization}

  Next, let us consider the case of multi-dimensional non-integrable
Hamiltonian system.  For such systems, as is well known, the
periodic-orbit quantization method provides us with a useful base
toward understanding the correspondence between classical periodic
orbits and properties of quantum
spectra.\refmk{\Ref{GutzLH},\Ref{Gutz}} This theory is essentially
based on the path integral formalism of quantum mechanics.  The first
step is to express the quantum level density
$g(E)=\sum_n\delta(E-E_n)$ as a trace of the energy-dependent
Green function;
\begin{eqnarray}
g(E)&=&-{1\over\pi}\Im\,\Tr{1\over E+i\epsilon-\hat{H}}
\nonumber \\
&=&-{1\over\pi}\Im\int\d\bbox{q}G(\bbox{q},\bbox{q};E)~.
\label{ldens}
\end{eqnarray}
The Green function $G(\bbox{q}'',\bbox{q}';E)$ is a Fourier
transform of the transition amplitude
$K(\bbox{q}'',t;\bbox{q}',0)=\bra{\bbox{q}''}\exp(-it\hat{H}/\hbar)
\ket{\bbox{q}'}\iprod\theta(t)$,
and we can express it in the path integral form.  Evaluation of the
path integral by the stationary phase approximation (SPA) extracts the
classical trajectories.  The Fourier transformation is also performed
by means of the SPA.  Finally, the trace integral appearing in
Eq.~\Eqn{ldens} extracts the periodic orbits and one obtains the
following expression called the Gutzwiller trace formula:
\begin{equation}
g(E)\simeq\bar{g}(E)+\sum_{n,\gamma}A_{n\gamma}(E)
  \cos\left(nS_\gamma(E)/\hbar-(\pi/2)\mu_{n\gamma}\right)~,
\label{tracef}
\end{equation}
where $\bar{g}(E)$ denotes the average level density and the second
term on the r.h.s. represents the oscillating part.  The summation is
taken over all periodic orbits and their multiple traversals.
$S_\gamma$ is a classical action along the orbit $\gamma$,~
$S_\gamma=\oint_\gamma\bbox{p}\iprod\d\bbox{q}$, and $\mu_\gamma$ is a
Maslov phase.  The amplitude factor $A_{n\gamma}$ depends on the phase
space structure about the periodic orbit $\gamma$, as we shall discuss
in \S2.4.  For sufficiently isolated orbits, the trace integral is
well approximated by the SPA and the amplitude factor for the $n$-fold
traversal of orbit $\gamma$ can be written as\refmk{\Ref{Gutz}}
\begin{equation}
A_{n\gamma}={1\over\pi\hbar}
{T_\gamma\over\sqrt{|\det(\Id-M_\gamma^n)|}}~,
\label{ampiso}
\end{equation}
where $T_\gamma$ and $M_\gamma$ represent the period and the monodromy
matrix of the primitive orbit $\gamma$, respectively.  This expression
is known to work well for chaotic systems such as
billiards.\refmk{\Ref{Gutztext}}

\subsection{Stability of Classical Trajectories}

  The amplitude factor in the trace formula is related with the
properties of phase space around the periodic orbits.  Let us write
the Hamilton equation in $2f$-dimensional phase space as
\begin{equation}
{\d\over\d t}Z=\Lambda\nabla H~,
\end{equation}
with
\begin{eqnarray}
Z={\bbox{p}\choose\bbox{q}},~~
\nabla={{\bbox{\nabla}_p}\choose{\bbox{\nabla}_q}},~~
\Lambda={{0~-1}\choose{1\hfill 0}}~,
\nonumber
\end{eqnarray}
and consider the time evolution of the deviation $\delta Z(t)$ from
the reference classical trajectory $Z_0(t)$.  To the first order in
$\delta Z$, we obtain
\begin{equation}
{\d\over\d t}\delta Z=\Lambda{\cal H}\delta Z~,
\label{jacobieq}
\end{equation}
where ${\cal H}$ is the Hessian matrix defined by
\begin{eqnarray}
{\cal H}={{H_{pp}~~H_{pq}}\choose{H_{qp}~~H_{qq}}}_{Z_0}~,
\hspace{8mm}
(H_{pq})_{ij}={\partial^2 H\over\partial p_i\partial q_j}
\hspace{8mm}\hbox{etc.}
\label{hessian}
\end{eqnarray}
Knowing ${\cal H}(t)$, Eq.~\Eqn{jacobieq} can be easily integrated;
\begin{eqnarray}
\delta Z(t)=\exp\left[
{\Lambda\int_0^t\d t'\,{\cal H}\big(Z_0(t')\big)}
\right]\delta Z(0)
\equiv{\cal M}(t)\,\delta Z(0)~.
\label{intjacobi}
\end{eqnarray}
${\cal M}$ is called the stability matrix. It is real and symplectic;~
$\Lambda{\cal M}^T\Lambda^{-1}={\cal M}^{-1}$.  When one take a
periodic orbit as the reference trajectory and the period $T$ as time
$t$, the stability matrix is particularly called monodromy matrix
$M_\gamma\equiv{\cal M}(T_\gamma)$.\refmk{\Ref{BarDav}} It is known
that the eigenvalues of $M_\gamma$ are independent of the initial
condition $Z(0)$ on the orbit.  According to the symplectic property
of the monodromy matrix, its eigenvalues appear in pairs
$(+/-)(e^\alpha, e^{-\alpha})$, where $\alpha$ is real or pure
imaginary.  When $\alpha$ is pure imaginary ($\alpha=iw$), the orbit
is stable and torus exists surrounding it. $w$ is called winding
number of the torus.  When $\alpha$ is real ($\alpha=\lambda T$), the
orbit is unstable and $\lambda$ is called the Lyapunov exponent which
measures the degree of instability.

\subsection{Magnitudes of Shell Effects}

  Strength of the shell effect depends mainly on three factors, to be
discussed below, associated with the periodic
orbits.\refmk{\Ref{StrMag}}

  The first factor is the degeneracy of the orbit.  Here the term
`degeneracy' means the number of independent continuous parameters
(additional to energy) that specify a certain orbit from continuous
family of orbits having the same action.  For example, planar orbits
in a spherical potential form continuous family generated by rotation
and the degeneracy is generally 3, since a certain orbit belonging to
this family is specified by three Euler angles.  As illustrated by the
above example, degeneracy is related to continuous symmetry of the
system.  These degeneracies correspond to the unit eigenvalues of $M$.

  The second factor is the stability of the orbit.  For non-integrable
systems, evaluating the trace integral by the SPA, one sees that the
amplitude behaves as
\begin{eqnarray}
A_{n\gamma}\propto{1\over\sqrt{|\det(\Id-\tilde{M}_\gamma^n)|}}~,
\label{ampstb}
\end{eqnarray}
where $\tilde{M}$ is a reduced monodromy matrix in which degrees of
freedom corresponding to the unit eigenvalues of $M$ are excluded out.
The more unstable is the orbit, the weaker is its contribution to the
level density, because it has a large Lyapunov exponent and the
denominator on the r.h.s. becomes large.  The above proportionality is
valid only when all the eigenvalues of $\tilde{M}$ are sufficiently
distant from unity.  However, one of the eigenvalues may happen to be
very close to unity.  This is called nonlinear resonance where two
frequencies of the torus coincide with each other and gives rise to a
periodic orbit bifurcation.  Namely, the period $n$-upling bifurcation
occurs when $\det(\Id-\tilde{M}_\gamma^n)=1$.  In this resonance
region, one have to use a more sophisticated treatment than the SPA;
for example, the uniform approximation using the resonant normal form.
Such a procedure is formally discussed by Ozorio de Almeida and
Hannay,\refmk{\Ref{OzoHan}} but, to the best of our knowledge,
application of this theory to multi-dimensional, non-integrable
Hamiltonian system has not been performed yet.

  The third factor is the phase space
volume occupied by an orbit.  It is not important in our analysis
because it is insensitive to the variation of potential parameters.

  Let us examine how these three factors enter in the amplitude
factors for different types of periodic orbits.  First, consider a
chaotic orbit, that is, a well-isolated orbit whose degeneracy equals
zero.  Its amplitude factor is given by Eq.~\Eqn{ampiso}.  There
appears the same stability factor as Eq.~\Eqn{ampstb}, and the period
$T$ measures the phase space volume of the orbit.  Next, as an example
of non-isolated orbits, let us consider orbits in axially symmetric
deformed potentials, whose degeneracy equals one corresponding to the
rotation about the symmetry axis.  It is convenient to use the
cylindrical coordinates $(\rho,\varphi,z)$.  One should then perform
the integral uniformly with respect to $\varphi$ in the trace, because
in this direction periodic orbits exist continuously.  Thus we obtain
the following expression of the amplitude factor for these orbits (see
Appendix A):
\begin{eqnarray}
A_{n\gamma}(E)
&=&{4\pi\over(2\pi\hbar)^{3/2}}
  {B_\gamma\over\sqrt{|2-\Tr\tilde{M}_\gamma^n|}}~,
\label{AXAMP} \\
B_\gamma
&=&\int_0^{T_\gamma}\d t
\left|{\partial\varphi(t+T_\gamma)\over\partial p_\varphi(t)}
\right|^{-1/2}.
\nonumber
\end{eqnarray}
Here, $B_\gamma$ contains the first and the third factors
mentioned above. The period $n$-upling bifurcation occurs when
$\Tr\tilde{M}_\gamma^n=2$.

\section{Supershell Effect in the SD Oscillator}

  In this section we apply the semiclassical theories to the
axially-symmetric 2:1 deformed harmonic oscillator Hamiltonian
\begin{equation}
H_0(\bbox{p},\bbox{q})
  ={~\bbox{p}^2\over 2M}+\sum_{i=x,y,z}{M\omega_i^2q_i^2\over 2}~,
\label{sdhamil}
\end{equation}
where $\omega_x=\omega_y\equiv\omega_\perp=2\omega_z\equiv2\osh$,
and we discuss how the supershell structure emerges in this case.  We
compare the two semiclassical quantization methods summarized in
the preceding section, and discuss their relations.

\subsection{The Periodic Orbit Method}

  In this subsection we analyze the supershell effect in the SD
oscillator defined by \Eqn{sdhamil} using the Gutzwiller trace
formula.  The trace formula can be derived also for such an integrable
system if the degeneracy (mentioned below) is properly taken into
account.  According to Ref.~$\Ref{Magner}$, the semiclassical level
density may be written as
\begin{equation}
g(E)\simeq\bar{g}(E)
   +\Sup(\gosc,{(IV)})(E)
   +\Sup(\gosc,{(II)})(E)~.
\label{sdld}
\end{equation}
The first term on the r.h.s. represents the mean level density;
\begin{equation}
\bar{g}(E)
={1\over(2\pi\hbar)^3}\int\d\bbox{p}\d\bbox{q}\,
  \delta\left(E-H(\bbox{p},\bbox{q})\right)
={E^2\over8(\hbar\osh)^3}~.
\label{tfld}
\end{equation}
The second and the third terms are the oscillating parts representing
the shell effects.  The superscript {\small (II)} and {\small (IV)}
denote the degeneracies of the periodic orbits.
$\Sup(\gosc,{(IV)})(E)$ is a contribution from four-hold
degenerate orbits whose periods are multiples of
$\Sup(T,{(IV)})=2\pi/\osh$.  The present model is very special in the
sense that all trajectories are periodic, and one should explicitly
perform four integrals (corresponding to the degeneracy) in evaluating
the trace formula.  Thus one obtains the following expression;
\begin{eqnarray}
\Sup(\gosc,{(IV)})(E)
&=&\sum_{m\ne 0}
  {~E^2\over8(\hbar\osh)^3}
  \cos\left[m\left({\Sup(S,{(IV)})\over\hbar}-(4+4+2){\pi\over2}
  \right)\right]
\nonumber \\
&=&\sum_{m=-\infty}^\infty
  {~E^2\over8(\hbar\osh)^3}
  \cos\left[m\left({E\Sup(T,{(IV)})\over\hbar}-5\pi
  \right)\right]-\bar{g}(E)
\label{tfpob} \\
&=&\sum_N{(N+5/2)^2\over8}
  \delta\left(E-\hbar\osh\left(N+{5\over2}\right)\right)
  -\bar{g}(E)~,
\label{tfldb}
\end{eqnarray}
where $\Sup(S,{(IV)})=E\Sup(T,{(IV)})$ is the action integral along
the primitive periodic orbit and the sum over $m$ accounts for
multiple traversals.  The last expression is obtained by using the
Poisson sum formula
\begin{equation}
\sum_{m=-\infty}^\infty \exp({2\pi imA})
  =\sum_{N=-\infty}^\infty \delta(A-N)~.
\label{sfpoisson}
\end{equation}
In Eq.~\Eqn{sdld}, $\Sup(\gosc,{(II)})(E)$ is the contribution from
two-fold degenerate orbits whose periods are odd integer times
$\Sup(T,{(II)})=\Sup(T,{(IV)})/2$.  It is obtained in a similar
manner, except that the integrations with respect to the $z$ direction
may be performed by the SPA.  The result is written as
\begin{eqnarray}
\Sup(\gosc,{(II)})(E)
&=&\sum_{m'={\rm odd}}
  {~E\over8(\hbar\osh)^2\sin(m'\pi/2)}
  \sin\left[m'\left({\Sup(S,{(II)})\over\hbar}-(2+2){\pi\over2}
  \right)\right]
\nonumber \\
&=&\sum_{m=-\infty}^\infty
  {~E\over8(\hbar\osh)^2}
  \cos\left[(2m+1)\left({E\Sup(T,{(II)})\over\hbar}-{5\pi\over2}
  \right)\right]
\label{tfpoe} \\
&=&\sum_N(-)^N{N+5/2\over8}
  \delta\left(E-\hbar\osh\left(N+{5\over2}\right)\right)~,
\label{tflde}
\end{eqnarray}
when the sum over $m'$ accounts for multiple traversals of the
primitive periodic orbit.  The expressions \Eqn{tfpob} and \Eqn{tfpoe}
were first derived in Ref.~$\Ref{Magner}$.

  Summing up the above three contributions,
we obtain the degeneracy $d_N$ of the $N$-th shell as
\begin{equation}
d_N={([N/2]+1)([N/2]+2)\over2}+{3\over32}~,
\end{equation}
where $[\ast]$ is the Gauss symbol.  The first term on the r.h.s.
correspond to the exact degeneracy of the quantum spectrum.  We thus
see that the result obtained by the trace formula is very accurate
(the deviation from the exact quantum result is only 3/32).

  Now let us focus our attention on a smoothed density of levels with
finite energy resolution $\delta E=\hbar\osh$.  It is then sufficient
to consider finite number of periodic orbits of short periods
satisfying the following uncertainty relation;
\begin{equation}
T\leq T_{\rm max}
={2\pi\hbar\over\delta E}~.
\end{equation}
As far as gross properties of the level density is concerned,
therefore, the well-known problem of the long time propagation in the
semiclassical approximation does not occur.

  We show in Figs.~\ref{SPSD}-a) and \ref{SPSD}-b) the contributions
from the families of periodic orbits with periods $2\pi/\osh$ and
$2\pi/\omega_\perp$, respectively, and in Fig.~\ref{SPSD}-c) the sum
of them.  There appears an undulating pattern in the level density due
to the interference of the above two families of periodic orbits,
which is just the supershell structure.  Thus, one sees that the
supershell structure emerges from this interference effect.

\subsection{The EBK Method}

  Defining the action-angle variables $(I_i,\theta_i)$ by
\begin{equation}
  \begin{array}{rl}
  p_i&=\sqrt{\mathstrut 2M\omega_iI_i}\sin\theta_i~, \\
  q_i&=\sqrt{\mathstrut 2I_i/M\omega_i}\cos\theta_i~,
  \end{array}
\end{equation}
the Hamiltonian~\Eqn{sdhamil} is written as a function of only action
variables as $H_0(\bbox{I})=\bbox{\omega}\iprod\bbox{I}$.  The
Maslov indices are 2 for all paths $i$, and therefore the EBK
quantization condition becomes
\begin{equation}
I_i=\hbar(n_i+1/2)~.
\end{equation}
In the present case, this EBK quantization gives exact quantum
eigenvalues: $E=\sum_i\hbar\omega_i(n_i+1/2)$.  Now let us investigate
the roles of classical periodic orbits in giving rise to the
supershell structure in the quantum spectrum.  For this purpose, we
use the method of topological sum.\refmk{\Ref{BerTab}} The
semiclassical level density is written as
\begin{equation}
g(E)=\sum_{\{n\}}\delta\Big(
  E-\sum_i\hbar\omega_i(n_i+\alpha_i/4)\Big)~.
\label{sdscld}
\end{equation}
Using the Poisson sum formula~\Eqn{sfpoisson}, one can rewrite
Eq.~\Eqn{sdscld} as
\begin{equation}
g(E)=g_0(E)+\sum_{\bbox{M}\ne0}g_{{}_M}(E)~,
\end{equation}
where
\begin{equation}
g_{{}_M}(E)={1\over\hbar^3}\int_0^\infty\d\bbox{I}\,
  \delta\big(E-\bbox{\omega}\iprod\bbox{I})\exp(2\pi i
  \bbox{M}\iprod(\bbox{I}/\hbar-\bbox{\alpha}/4)\big)~,
\end{equation}
and the summation is taken over all the combinations of integers,
$\bbox{M}=(M_x,M_y,M_z)$.  Here, $g_0(E)$ represents a mean level
density corresponding to the Thomas-Fermi approximation;
\begin{equation}
g_0(E)={1\over\hbar^3}\int_0^\infty\d\bbox{I}\,
  \delta(E-\bbox{\omega}\iprod\bbox{I})
  ={E^2\over8(\hbar\osh)^3}~.
\end{equation}
The remaining terms with non-zero \bbox{M} represent the oscillating
part responsible for the shell structure.  To simplify the expression,
we introduce the notation $f_i=2\pi M_i\osh/\omega_i$.  The dominant
contribution comes from terms satisfying the resonance condition
$\bbox{M}=\bbox{M}^\ast\propto\bbox{\omega}$, \ie, $f_x=f_y=f_z$
(in the present case, $\bbox{M}=m(2,2,1)$).
Carrying out the integration with respect to \bbox{I} and denoting the
sum over such resonant terms as $\Sup(\gosc,{(i)})$, we obtain
\begin{equation}
\Sup(\gosc,{(i)})(E)={E^2\over8(\hbar\osh)^3}
  \sum_{m\ne0}e^{2\pi im(E/\hbar\osh-5/2)}~.
\label{ebkpob}
\end{equation}
Next, let us consider the `partially resonant' terms which satisfy the
condition~ $f_i=f_j\equiv f_\perp\ne f_k$.
Carrying out the integration with respect to \bbox{I}, they are
evaluated as
\begin{eqnarray}
g&&\!{}_{{}_{M({\rm part.res.})}}(E)
  \nonumber \\
&&={1\over4(\hbar\osh)^2}\left(
  {E\over i(f_\perp-f_k)}e^{if_\perp E/\hbar\osh}
  +{\hbar\osh\over(f_\perp-f_k)^2}
  (e^{if_\perp E/\hbar\osh}-e^{if_kE/\hbar\osh})\right)
  e^{i\pi M\cdot\alpha/2}
  \nonumber \\
&&\simeq{1\over4(\hbar\osh)^2}\,
  {E\over i(f_\perp-f_k)}e^{if_\perp E/\hbar\osh}
  e^{i\pi M\cdot\alpha/2}~,
\end{eqnarray}
where the second term on the r.h.s. is neglected because it is
higher-order in $\hbar$.
Let us then take terms with $f_x=f_y\equiv f_\perp\ne f_z$, and write
\bbox{M} as $(m',m',l)$. Summing over terms with odd-$m'$ and
arbitrary integer $l$, we obtain
\begin{eqnarray}
\Sup(\gosc,{(ii)})(E)
&=&\sum_{m'={\rm odd}}\sum_{l=-\infty}^\infty
  {1\over4(\hbar\osh)^2}{E\over 2\pi i(m'/2-l)}
  e^{2\pi im'E/2\hbar\osh}\iprod e^{-2\pi i\{2m'+l\}/2}
\nonumber \\
&=&{E\over8(\hbar\osh)^2}
  \sum_{m=-\infty}^\infty e^{i\pi(2m+1)(E/\hbar\osh-5/2)}~.
\label{ebkpoe}
\end{eqnarray}
The last expression is obtained using the expansion formula of
cosecond in partial fractions:
\begin{equation}
{\rm cosec}\,z=\sum_{l=-\infty}^\infty {(-)^l\over z-l\pi}~.
\end{equation}
It can be easily shown in a similar way that the sum of other terms
which are the same order in $\hbar$ as \Eqn{ebkpoe} vanishes.

\subsection{Relation between the Two Methods}

  Now, let us discuss the correspondence between the results obtained
by the periodic-orbit method and the EBK method.
It is evident that the two terms
$\Sup(\gosc,{(i)})(E)$ and $\Sup(\gosc,{(ii)})(E)$
in the EBK treatment are identical with the contributions
$\Sup(\gosc,{(IV)})(E)$ and $\Sup(\gosc,{(II)})(E)$
which are evaluated by the trace formula for the periodic orbits with
periods $2\pi/\osh$ and $2\pi/\omega_\perp$, respectively.  This
result is very instructive to understand the physical meanings of the
resummation with respect to the indices $\bbox{M}$ by the use of the
Poisson sum formula.  The indices satisfying the resonance condition,
$\bbox{M}=m(2,2,1)$, correspond to a family of classical orbits with
periods $2m\pi/\osh$.  One can examine this by comparing
Eqs.~\Eqn{tfpob} and \Eqn{ebkpob}.  On the other hand, for
partially-resonant contributions, we find the correspondence in the
following way.  Comparing Eqs.~\Eqn{tfpoe} and
\Eqn{ebkpoe}, we notice that the family of planar orbits in the
$(x,y)$ plane corresponds to the summation over indices
$\bbox{M}=(1,1,l)$ with $-\infty<l<\infty$.  Its $m$-fold
traversals are related with $\bbox{M}=(m,m,l)$.  These
partially-resonant terms play
important roles in formation of the supershell structure in the
present model.

\section{Reflection-Asymmetric SD Oscillator Model}

\subsection{Model Hamiltonian}

  Let us consider a model Hamiltonian consisting of an
axially-symmetric 2:1 deformed harmonic oscillator and a
doubly-stretched octupole $(Y_{30})$ deformed potential;
\begin{equation}
H={~\bbox{p}^2\over2M}+M\omega_0^2\left(
{~r^2\over2}-\loct\,r^2\,Y_{30}(\theta)\right)'',
\label{chamil}
\end{equation}
where double primes indicate that the variables in parenthesis are
defined in terms of the doubly-stretched coordinates
$q_i''=(\omega_i/\omega_0)q_i$,~ and
$\omega_0=(\omega_x\omega_y\omega_z)^{1/3}$.  As emphasized by
Sakamoto and Kishimoto\refmk{\Ref{SakKis}}, the doubly-stretched
coordinates are suited to description of systems having quadrupole
equilibrium deformations, and possess several advantages over the
usual coordinates; for example, the center of mass motion is exactly
decoupled from the octupole-type deformations described by the above
Hamiltonian.  Note that the doubly-stretched octupole operator is in
fact a linear combination of the ordinary dipole and octupole
operators, although we sometimes omit the adjective
``doubly-stretched'' for simplicity.  In \Eqn{chamil}, we adopt the
quadratic radial dependence for the octupole-deformed potential for
the ease of taking into account the volume conservation condition.  By
requiring the volume surrounded by an equipotential surface to be
independent of the octupole deformation parameter $\loct$, the
$\loct$-dependence of $\omega_0$ is determined as
\begin{equation}
\omega_0(\loct)=\omega_0(0)\left[{1\over4\pi}\int\d\Omega
  (1-2\loct Y_{30}(\Omega))^{-3/2}\right]^{1/3}.
\label{vcons}
\end{equation}
We note that the average level density $\bar{g}(E)$ is independent of
$\loct$ when $\omega_0$ satisfies Eq.~\Eqn{vcons}.  Let us define
dimensionless variables as
\begin{equation}
\left\{~
  \begin{array}{rl}
    p_i & \rightarrow~ \sqrt{\mathstrut M\hbar\omega_0}~p_i~, \\
    q_i & \rightarrow~ \sqrt{\mathstrut \hbar/M\omega_0}~q_i~, \\
    H   & \rightarrow~ \hbar\omega_0~H~.
  \end{array}
\right.
\label{ndim}
\end{equation}
Then the Hamiltonian~\Eqn{chamil} becomes
\begin{equation}
H={~\bbox{p}^2\over2}+\left({~r^2\over2}-\loct\,r^2\,
    Y_{30}(\theta)\right)''.
\label{shamil}
\end{equation}
Since this potential is a homogeneous function of the second order
in coordinates, the scaling relation
\begin{equation}
H(\alpha\bbox{p}, \alpha\bbox{q})=\alpha^2H(\bbox{p}, \bbox{q})
\label{scaling}
\end{equation}
holds.  Thus, if $(\bbox{p}(t), \bbox{q}(t) )$ is a solution of the
Hamilton equation with energy $E$, $(\alpha\bbox{p}(t),
\alpha\bbox{q}(t) )$ is also a solution but with energy
$\alpha^2E$.  Namely, once the classical properties of the system are
known on a certain energy surface $E_0$, properties on other energy
surface $E$ are obtained by scale transforming the phase space
variables $~(\bbox{p},\bbox{q})~$ to $(\alpha\bbox{p},\alpha\bbox{q})$
with $\alpha$=$\sqrt{E/E_0}$.

\subsection{Supershell Structure}

  Figure~\ref{OLD} shows the oscillating part of the level density for
the Hamiltonian~\Eqn{shamil} calculated by means of the Strutinsky
method.  A characteristic property of the oscillating level density is
that it exhibit the supershell pattern.  Figure~\ref{SUPS} gives a
phenomenological illustration of the concept of the supershell.  It is
seen from this figure that the oscillating level density can be
represented as a superposition of trigonometrical functions,
$\cos(ET_\gamma/\hbar)$ with $T_\gamma\approx2\pi/\omega_\perp$ and
$2\pi/\osh$, in a good approximation.  We shall confirm in \S6 that
this supershell pattern in fact arises from an interference effect
between two families of classical periodic orbits with periods
$T_\gamma\approx2\pi/\osh$ and $2\pi/\omega_\perp$.

  It should be recalled here that the important factor from the point
of view of gaining the shell-structure energy is not the heights of the
maxima but the depths of the minima in the oscillating level density.
Needless to say, the minima in Fig.~\ref{SUPS} correspond to the
closed-shell configurations with respect to the SD major shell quantum
number $\Nsh$.  Let us notice how the depths of the minima change as
functions of $\loct$.  Then we find that the minima associated with the
odd-$\Nsh$ closures becomes shallower as $\loct$ increases, whereas
those at the even-$\Nsh$ closures are tough.  Consequently, the
odd-even staggering of the minima with respect to the $\Nsh$ quantum
number develops with increasing $\loct$.
Possible mechanisms of the enhancement of the supershell structure
will be discussed in \S6.

\section{Classical Analysis}

  In this section, we discuss the classical-mechanical properties of
the single-particle motion in the reflection-asymmetric SD potential
defined in the preceding section.

\subsection{Poincar\'e Map}

  Let us examine classical phase space structure by plotting the
Poincar\'e map.\refmk{\Ref{Gutztext}}  Since our Hamiltonian is
axially symmetric, it reduces to a two-dimensional one with the
cylindrical coordinates $(\rho,z)$ and with a definite angular
momentum $p_\varphi=m$;
\begin{equation}
H={1\over2}\left(p_\rho^2+p_z^2\right)
+V_{\rm eff}(\rho,z;m)~,
\end{equation}
where
\begin{equation}
V_{\rm eff}(z,\rho;p_\varphi)
={~p_\varphi^2\over2\rho^2}+{4\rho^2+z^2\over2}-\loct
\sqrt{{7\over4\pi}}\,{z^3-6z\rho^2\over\sqrt{4\rho^2+z^2}}~.
\end{equation}
We can examine the Poincar\'e map for each value of $m$.  It is
convenient to choose the Poincar\'e section $\Sigma$ as the surface
with $p_\rho=0$, which is intersected by any trajectory.
Figure~\ref{PMAPA} shows calculated Poincar\'e maps $(z, p_z)$ for the
Hamiltonian~\Eqn{chamil} with various values of the octupole
deformation parameter $\loct$.  We see that the system is
quasi-integrable for small $\loct$ and almost all the phase space is
foliated by KAM tori.  With increasing $\loct$, however,
chaotic regions begin to spread out from the hyperbolic points.
Figure~\ref{PMAPC} shows Poincar\'e maps for different values of
$p_\varphi$.  We note that the phase space volume corresponding to the
$(\rho, z)$ degrees of freedom contract and the system becomes
more regular as $p_\varphi$ increases.

  Figure~\ref{PMAPB} shows Poincar\'e maps for the surface of section
$(\rho, p_\rho)$ with $z=0$.
The origin corresponds to the linear orbit along the $z$-axis and the
structure around it is exhibited.

\subsection{Periodic Orbits and Their Bifurcations}

  While all trajectories are periodic when $\loct=0$, only very
limited trajectories remain as periodic orbits when $\loct\ne0$.  In
the Poincar\'e section, centers of tori correspond to stable periodic
orbits, while saddles to unstable ones.  We calculate the periodic
orbits by the monodromy method proposed by Baranger, Davies and
Mahoney.\refmk{\Ref{BarDav}} In this method, periodic orbits are found
in an iterative manner starting with approximate closed curves. By
gradually changing $\loct$, we can use the periodic orbits found for a
slightly smaller value of $\loct$ as inputs for this procedure.
Figure~\ref{POC} shows short periodic orbits for
Hamiltonian~\Eqn{chamil} with $\loct=0.4$ obtained in this way.  Also
shown in Fig.~\ref{POP} are planar orbits for $\loct=0.3\sim0.4$.
As $\loct$ increases, the phase space structure becomes more
complicated due to bifurcations of stable periodic orbits.  For
example, a period-tripling bifurcation of orbit A occurs at
$\loct\simeq0.36$.  Thereafter, orbit A bifurcates into orbits 3A
(triple traversal of orbit A), E and F.
In the Poincar\'e map (see Fig.~\ref{PMAPA}) for $\loct=0.37$, we
can see three resonant island chains surrounding the central KAM
torus, which are associated with the newly-born periodic orbits E
and F.
Likewise, a period-doubling bifurcation of orbit B occurs at
$\loct\simeq0.4$, from where orbit B bifurcates into orbits 2B (double
traversal of orbit B) and K.  Many higher order bifurcations occur
almost everywhere in regular regions of the phase space.  Properties
of the calculated periodic orbits are summarized in Table~I for
several values of $\loct$.

\section{Semiclassical Analysis}

\subsection{The Cause of the Enhancement of the Supershell Effect}

  As mentioned in \S2.4, strength of shell effect associated with a
periodic orbit is mainly determined by degeneracy and stability of the
orbit.  Let us discuss how these properties change when the octupole
deformation is added to the SD oscillator.  When $\loct=0$, orbits
with the period $2\pi/\osh$ and those with the period
$2\pi/\omega_\perp$ have different degeneracies, 4 and 2,
respectively.  Therefore, the shell effect originating from the former
families of orbits is much stronger than that from the latter.  As a
result, the interference effect between the two families of periodic
orbit, \ie, the supershell effect, is rather weak.  When $\loct\ne0$,
in general, degeneracies of the orbits reduce to 1. (For orbits D and
A' in Fig.~\ref{POC} having special symmetry, the degeneracy is 0.)
Thus, generally speaking, shell effects at $\loct\ne0$ are expected to
become weaker in comparison with those in the $\loct=0$ limit, and it
seems hard to understand the enhancement mechanism of the supershell
effect at $\loct\ne0$ in terms of the degeneracy property.

  Next, let us consider the other factor, \ie, stability of periodic
orbits.  In Fig.~\ref{TRM} we show calculated values of $\Tr M$ for
relevant orbits as functions of $\loct$.  At $\loct=0$, orbits with
the period $2\pi/\osh$ are resonant ($\Tr M=2$), while orbits with the
period $2\pi/\omega_\perp$ are non-resonant and take $\Tr M=-2$.  With
increasing $\loct$, $\Tr M$ for orbits 2A(double traversal of orbit A)
and B decrease and deviate from 2.  At $\loct\simeq0.4$, a
period-doubling bifurcation of orbit B occurs; a new stable orbit K is
created and orbit B becomes unstable.  Orbits C and C' are unstable
for $\loct>0$ and their values of $\Tr M$ become larger as $\loct$
increases.  According to the argument in \S2.4, we thus expect that
the contributions of these orbits to the shell effect decrease with
increasing $\loct$.  On the other hand, orbit A is stable and its $\Tr
M$ value approaches towards 2 as $\loct$ increases.  This implies that
the contribution of orbit A becomes more important.  In this way,
relative magnitude of the amplitude factors between the two families
of orbits with the period $\approx2\pi/\osh$ and
$\approx2\pi/\omega_\perp$ changes so that the interference effect
between them becomes stronger.

  The above discussion is based on the expression~\Eqn{AXAMP} obtained
by the SPA.  We should note, however, that our classical phase space
contains both regular and chaotic regions, \ie, our system is a mixed
system.  As is well known, such a system is abundant in the resonance
regions where the SPA breaks down, so that the amplitude factors
$A_{n\gamma}$ should be evaluated by means of a more sophisticated
method beyond the SPA, \eg, the uniform
approximation.\refmk{\Ref{OzoHan}} This is an interesting future
subject, and we expect that the above consideration will remain valid,
as long as a qualitative feature is concerned, even when nonlinear
effects beyond the SPA are taken into account.

\subsection{Scaling Properties and Fourier Analysis}

  By virtue of the scaling property, Eq.~\Eqn{scaling}, the following
scaling rules hold;
\begin{equation}
\left\{~~
  \begin{array}{rl}
    S_\gamma(E) & =E T_\gamma~, \\
    \bar{g}(E)  & =E^2\bar{g}(1)~, \\
    A_\gamma(E) & =E^{d_\gamma/2}A_\gamma(1)~,
  \end{array}
\right.
\label{scalerule}
\end{equation}
where $d_\gamma$ denotes the degeneracy of orbit $\gamma$;
$d_\gamma=1$ for a general orbit and 0 for an isolated orbit
like D or A' in Fig.~\ref{POC}.

  Let us consider the Fourier transform
\begin{equation}
P(s)=\int\d E\,e^{isE}E^{-d/2}g(E)
\label{powers}
\end{equation}
of the level density $g(E)$ multiplied by $E^{-d/2}$.  (The factor
$E^{-d/2}$ is attached here to compensate for the energy dependence of
the amplitude factor $A_\gamma$; see below.)  If one insert the exact
level density $g(E)=\sum_n\delta(E-E_n)$, it becomes
\begin{equation}
P^{\rm(qm)}(s)=\sum_n E_n^{-d/2}e^{isE_n}
\label{qpower}
\end{equation}
This quantity can be evaluated with the use of the eigenvalues
obtained by a quantum mechanical calculation.  On the other hand,
if we insert the semiclassical level density \Eqn{tracef} in
\Eqn{powers} and put $d=1$ appropriate to non-isolated orbits, then
we obtain
\begin{equation}
P^{\rm(cl)}(s)=\bar{P}(s)
  +\sum_{n,\gamma}A_{n\gamma}(1)e^{i\pi\mu_{n\gamma}/2}
  \delta(s-nT_\gamma)~.
\label{clpower}
\end{equation}
Here $\bar{P}(s)$ comes from $\bar{g}(E)$ and has a peak at $s=0$
associated with the orbits of zero length.  On the other hand, the
second term on the r.h.s. gives rise to sharp peaks at $s=nT_\gamma$
associated with the classical periodic orbits $\gamma$ with periods
$T_\gamma$ (and their multiple traversals).  Note that, owing to the
scaling property \Eqn{scalerule}, periods $T_\gamma$ of the primitive
orbits are equal to action $S_\gamma(1)$ calculated at $E=1$.  If the
trace formula is valid, one expect $P^{\rm(cl)}\simeq P^{\rm(qm)}$.
Thus, we can extract information about classical periodic orbits by
calculating $P^{\rm(qm)}$.  Namely, the amplitude factors and the
Maslov phases of the periodic orbits may be obtained from absolute
values and arguments of $P^{\rm(qm)}(s)$,
respectively.\refmk{\Ref{Balian},\Ref{Wintgen}}

  Now, let us evaluate the Fourier transform \Eqn{qpower}.  Since the
summation is taken over a finite number of quantum levels in practice,
we introduce the Gaussian cut-off and define a smoothed version of it;
\begin{equation}
P_{\Delta s}(s)=\int\d s'P(s')\,f((s'-s)/\Delta s)
\end{equation}
where $f(x)$ is Gaussian $f(x)=\exp(-x^2/2)$.  For ~\Eqn{qpower}
and \Eqn{clpower}, we obtain
\begin{eqnarray}
P_{\Delta s}^{\rm(qm)}(s)
&=&\sum_n E_n^{-d/2}e^{isE_n}f(E_n/E_{\rm max})~,
\label{avqpower} \\
P_{\Delta s}^{\rm(cl)}(s)
&=&\bar{P}_{\Delta s}(s)+\sum_{n,\gamma}
  A_{n,\gamma}(1)e^{i\pi\mu_{n\gamma}}
  f((s-nT_\gamma)/\Delta s)~,
\end{eqnarray}
where $E_{\rm max}=1/\Delta s$.

  We calculate the eigenvalues by a matrix diagonalization method with
the deformed oscillator bases, and use the lower part of the resulting
spectrum.  Figure~\ref{FTLA} shows the absolute value of $P_{\Delta
s}^{\rm(qm)}(s)$ for $\loct=0.2\sim0.4$ calculated with $E_{\rm
max}=15\hbar\osh(\loct)$.  The loci of the periods of classical
periodic orbits and their multiple traversals are indicated by arrows
in the figures.  We see nice correspondence between the peaks of
$P(s)$ and the periods of classical periodic orbits.  Almost all peaks
can be explained in terms of the classical orbits, indicating that the
properties of quantum spectrum is characterized mostly by classical
periodic orbits.

  Next, let us notice the $\loct$ dependence of
$P^{\rm(qm)}(s)$.  In Fig.~\ref{FTLA} we see that the peak at
$s\simeq2\pi/\osh$ decreases, while the peak at $s\simeq\pi/\osh$
grows up with increasing $\loct$.  Since heights of the peaks in
$P^{\rm(qm)}(s)$ indicate intensities coming from the
corresponding periodic orbits, this implies that the contributions
from the orbits with the period $\approx2\pi/\osh$ becomes
increasingly important as $\loct$ increases.  The change in relative
intensity as a function of $\loct$ between the two families of
periodic orbit seen in Fig.~\ref{FTLA} may be responsible for the
enhancement of the supershell effect in the reflection-asymmetric
SD potential, in accordance with our discussion in the preceding
subsection.

\subsection{Quantum Signature of Bifurcations}

  In order to see how the bifurcations (resonances) of periodic orbits
affect the magnitudes of the Fourier amplitudes, let us evaluate the
heights of the peak as functions of $\loct$ at the periods of the
classical orbits.  As examples, we take the period-tripling and the
period-5-upling bifurcations of orbit A, which occur at $\loct=0.36$
and 0.25, respectively.  In Fig.~\ref{FTO} are plotted the calculated
values of $\bar{P}^{\rm(qm)}(s)$ as functions of $\loct$ at specific
values of $s$ that correspond to the periods of three- and five-fold
traversals of orbit A.  In accordance with the argument given below
Eq.~\Eqn{ampstb}, we find that the heights of the peaks indeed exhibit
supremes near the bifurcation (resonance) points, but little delays
are observed for both cases.  To account for this delay, it may be
necessary to go beyond the SPA.

\subsection{Angular Momentum Decomposition of the Trace Formula}

  As our system is axially symmetric, the angular momentum about the
symmetry axis $p_\varphi$ is a good quantum number.  Thus, the
level density can be decomposed as
$g(E)=\sum_m g(E;m)$ with $m$ denoting the angular momentum quantum
number.
Let us derive a semiclassical
expression of $g(E;m)$.  Writing the three-dimensional
coordinate vector as $\bbox{q}=(\bbox{Q},\varphi)$ with
$\bbox{Q}=(\rho,z)$, the Green function may be decomposed as
\begin{eqnarray}
G(\bbox{q}'',\bbox{q}';E)
&=& \sum_{n=-\infty}^\infty G((\bbox{Q}'',\psi+2n\pi),(\bbox{Q}',0);E)
\nonumber \\
&=& \sum_n \int\d M e^{i(\psi+2n\pi)M}
    \tilde{G}(\bbox{Q}'',\bbox{Q}';E,M)
\nonumber \\
&=& \sum_{m=-\infty}^\infty e^{im\psi}
    \tilde{G}(\bbox{Q}'',\bbox{Q}';E,m)~,
\label{apgreen}
\end{eqnarray}
where $\tilde{G}$ denotes a Fourier transform of $G$ with respect to
$\psi=\varphi'-\varphi$, and where the Poisson sum formula is used in
obtaining the last expression.  Taking the trace of Eq.~\Eqn{apgreen},
one can derive the trace formula for $g(E;m)$ in a way similar to
\Eqn{tracef};\refmk{\Ref{MagKolStr}}
\begin{eqnarray}
g(E;m)
&=& -2\,\Im\int\d\bbox{Q}\rho\tilde{G}(\bbox{Q},\bbox{Q};E,m)
\nonumber \\
&\simeq& \bar{g}(E;m)
  +{1\over \pi\hbar}\sum_{n,\alpha}
  {\tau_\alpha\over \sqrt{|W_{n\alpha}|}}
  \cos(n\sigma_\alpha(E)/\hbar-(\pi/2)\mu_{n\alpha})~,
\label{aptracef}
\end{eqnarray}
where $\sigma_\alpha$ denotes the action integrals along the two
dimensional closed orbits $\alpha$ in the $(\rho,z)$ plane,
$\tau_\alpha$ the periods, and $\mu_\alpha$ the Maslov phases.  Using
the symplectic property of the monodromy matrix, $W_{n\alpha}$ can be
written as
\begin{equation}
W_{n\alpha}=\det\,(\Id-M_\alpha^n)
=2-\Tr\,(M_\alpha^n)~,
\label{Wvalue}
\end{equation}
where $M_\alpha$ is a 2$\times$2 monodromy matrix.  Note that, for
symmetric self-retracing orbits, $M_\alpha^n$ is different from
$\tilde{M}_\alpha^n$ appearing in Eq.~\Eqn{AXAMP}.  It is easily
seen that, due to the reflection symmetry with respect to the
$z$-axis, these orbits in the $(\rho, z)$ plane have periods half of
those in the three dimensional space.

  Now, for $m=0$, a scaling property holds so that we can use the
Fourier transformation technique.  Since the degeneracy of the orbits
is zero in the two-dimensional space, we put $d=0$ in
Eq.~\Eqn{avqpower}.  From the above consideration, one expect that the
Fourier transform will exhibit peaks, in addition to those
corresponding to the periods of closed orbits, also at half of the
periods of the three-dimensional symmetric self-retracing orbits.  The
results of calculation is shown in Fig.~\ref{FTLZ}.  Again we find a
clear correspondence between peaks of the Fourier transform and
periods of classical orbits.  As expected, peaks appear also at half
integer times the period of orbit A.

\section{Concluding Remarks}

  We have found a clear correspondence between the shell structure,
\ie, the oscillatory structure in the smoothed level density, and the
classical periodic orbits for single-particle motions in a
reflection-asymmetric SD oscillator potential.  We have then shown
that the supershell effect, \ie, an interference effect between two
families of the periodic orbits having periods approximately
$2\pi/\omega_\perp$ and $2\pi/\osh$, develops when the
reflection-asymmetric deformation increases.  This supershell effect
is in clear correspondence with the odd-even effect in $\Nsh$ pointed
out in Refs.~$\Ref{NazDob}$ and $\Ref{AriMat}$.  Possible origins of
this enhancement phenomena have been pointed out in connection with
stabilities of the classical periodic orbits.  Quantum signature of
the period-tripling bifurcation of the shortest-period orbit is also
pointed out.

  It should be emphasized that our model Hamiltonian system is a mixed
system where chaos and tori are intermixed; accordingly,
period-multipling bifurcations occur, as we have seen, rather
frequently when the reflection-asymmetric deformation parameter is
varied.  As is well known, the SPA breaks down at the bifurcation
points so that we cannot use the Gutzwiller trace formula for the aim
of calculating the smoothed level density.  Instead, by virtue of the
scaling property of our model Hamiltonian, we have been able to use
the Fourier transformation technique to find the quantum-classical
correspondence.  Properties of the Gutzwiller amplitudes have been
used only as a guide to qualitative discussions.  It is an interesting
future subject to investigate the problem discussed in this paper by
using a more sophisticated method, like the uniform approximation,
which goes beyond the SPA.
\bigskip

\centerline{\bf Acknowledgements}
\medskip

We thank
W.~Nazarewicz,
M.~Toda,
Y.~R.~Shimizu,
M.~Matsuo,
H.~Aiba,
S.~Mizutori
and
T.~Nakatsukasa
for useful discussions.

\newpage
\appendix
\section{Derivation of Eq.~{\protect\rm(\protect\ref{AXAMP})}}

  Semiclassical expression of the Green function in a
$f$-dimensional system is
\begin{equation}
G(\bbox{q}'',\bbox{q}';E)
=\sum{2\pi\over(2\pi i\hbar)^{(f+1)/2}}
\sqrt{|D_S|}\exp\biggl({i\over\hbar}S(\bbox{q}'',\bbox{q}';E)
-i\kappa{\pi\over2}\biggr),
\end{equation}
where $S(\bbox{q}'',\bbox{q}';E)$ is the classical action
$\int\bbox{p}\iprod\d\bbox{q}$ along a trajectory connecting
$\bbox{q}'$ and $\bbox{q}''$ with energy $E$.  The determinant $D_S$
in the amplitude factor is given by
\begin{equation}
D_S=\left|
  \begin{array}{cc}
  \partial^2S/\partial\bbox{q}''\partial\bbox{q}'
  &\partial^2S/\partial E\partial\bbox{q}' \\
  \partial^2S/\partial\bbox{q}''\partial E
  &\partial^2S/\partial E^2
  \end{array}
\right|
\label{sdet}
\end{equation}
Let us consider a three-dimensional system ($f$=3) with axial
symmetry, and define an orthogonal coordinate
$\bbox{q}=(\xi,\eta,\zeta)$ for each periodic orbit.  We take $\xi$
along the direction of the trajectory and $\eta$ perpendicular to both
$\xi$ and the azimuthal direction $\varphi$.  Differentiating the
Hamilton-Jacobi equations
\begin{mathletters}
  \begin{eqnarray}
  H(\bbox{p}''=\partial S/\partial\bbox{q}'',~\bbox{q}'')&=&E~,
  \label{hjeqa} \\
  H(\bbox{p}'=-\partial S/\partial\bbox{q}',~\bbox{q}')&=&E~,
  \label{hjeqb}
  \end{eqnarray}
\end{mathletters}
with respect to $E$ and using $\dot{\eta}=\dot{\zeta}=0$, one obtains
\begin{mathletters}
  \begin{eqnarray}
  1&=&\sum_i{\partial H\over\partial p''_i}
    {\partial^2S\over\partial q''_i\partial E}
    =\dot{\xi}''S_{\xi''E}~, \\
  1&=&-\sum_i{\partial H\over\partial p'_i}
    {\partial^2S\over\partial E\partial q'_i}
    =-\dot{\xi}'S_{E\xi'}~,
  \end{eqnarray}
\end{mathletters}
where $S_{xy}$ denotes $(\partial^2S/\partial x\partial y)$.  If one
differentiates \Eqn{hjeqa} and \Eqn{hjeqb} with $\xi'$ and $\xi''$,
respectively, one obtains
\begin{equation}
S_{q''_i\xi'}=S_{\xi''q'_i}=0
\end{equation}
Thus, the determinant \Eqn{sdet} is written as
\begin{equation}
D_S=\left|
  \begin{array}{cccc}
  S_{EE}    &S_{E\xi'} &S_{E\eta'}    &S_{E\zeta'}      \\
  S_{\xi''E} &0       &0              &0                \\
  S_{\eta''E} &0     &S_{\eta''\eta'} &S_{\eta''\zeta'} \\
  S_{\zeta''E} &0   &S_{\zeta''\eta'} &S_{\zeta''\zeta'}
  \end{array}
\right|
=-{1\over \dot{\xi}''}\,{1\over\dot{\xi}'} \left|
{{S_{\eta''\eta' }~~S_{\eta''\zeta' }}\atop
 {S_{\zeta''\eta'}~~S_{\zeta''\zeta'}}}
\right|~.
\end{equation}
If we use coordinates $(\xi,\eta,\varphi)$
which are generally not orthogonal, then we obtain
\begin{equation}
D_S={1\over J''\dot{\xi}''}{1\over J'\dot{\xi}'}\left|
{{S_{\eta''\eta'}~~S_{\eta''\psi}}\atop
 {S_{\psi\eta'  }~~S_{\psi\psi }}}
\right| ,
\end{equation}
where $J$ is Jacobian of the coordinate transformation, $J'$ and $J''$
denoting its value at $\bbox{q}'$ and $\bbox{q}''$, respectively, and
$\psi=\varphi''-\varphi'$.  Let us evaluate the trace of the
Green function in the stationary phase approximation.  As usual, the
action integral along a closed path may be expanded about a
stationary point $\bar{\bbox{q}}$ as
\begin{eqnarray}
S(\bbox{q},\bbox{q};E)
&=&S(\bar{\bbox{q}},\bar{\bbox{q}};E)
  +(\bbox{q}-\bar{\bbox{q}})^T\left[
   {\partial S\over\partial\bbox{q}''}
  +{\partial S\over\partial\bbox{q}'}\right]_{q''=q'=\bar{q}}
\nonumber \\
&&+{1\over 2}(\bbox{q}-\bar{\bbox{q}})^T \left[
  {\partial^2S\over\partial\bbox{q}''\partial\bbox{q}''}
  +2{\partial^2S\over\partial\bbox{q}''\partial\bbox{q}'}
  +{\partial^2S\over\partial\bbox{q}'\partial\bbox{q}'}
  \right]_{q''=q'=\bar{q}}(\bbox{q}-\bar{\bbox{q}})+\cdots
\label{expofs}
\end{eqnarray}
The stationary phase condition requires the second term in the r.h.s. to
vanish.  This is nothing but the condition for the
trajectory to be periodic, \ie, $\bbox{p}''=\bbox{p}'$.
Taking the axial symmetry into account, we can rewrite
Eq.~\Eqn{expofs} as
\begin{equation}
S(\bbox{q},\bbox{q};E)
=\bar{S}(E)+{1\over 2}W(\xi)\eta^2+\cdots~,
\end{equation}
where
\begin{equation}
W(\xi)=\det(\Id-M)S_{\eta''\eta'}
=(2-\Tr M)S_{\eta''\eta'}~.
\end{equation}
$M$ being the $(2\times2)$ monodromy matrix for the periodic orbit (see \S2.2).
Performing the Gauss-Fresnel integral with respect to $\eta$,
we finally obtain the following result
\begin{eqnarray}
g_{\rm osc}(E)
&=&{1\over 2\pi^2\hbar^2}\Im\sum \int\d\varphi\d\xi\d\eta J
  \sqrt{|D_S|}\exp\left[
  {i\over\hbar}\bigl(\bar{S}+{1\over 2}W(\xi)\eta^2\bigr)
  -i\kappa{\pi\over 2}\right]
\nonumber \\
&=&{4\pi\over(2\pi\hbar)^{3/2}}\sum_\gamma
  {B_\gamma\over\sqrt{|2-\Tr M_\gamma|}}
  \cos(S_\gamma/\hbar-\mu_\gamma\pi/2)~,
\end{eqnarray}
where  $\mu_\gamma=\kappa_\gamma-{\rm sign}(W_\gamma)/2$ and
\begin{eqnarray}
B_\gamma
&=&\oint_\gamma{\d\xi\over|\dot{\xi}|}\left|\left(\left|
  {S_{\eta''\eta'}~~S_{\eta''\psi} \atop
  S_{\psi\eta'} ~~S_{\psi\psi}}
\right|/S_{\eta''\eta'}\right)\right|^{1/2}
\nonumber \\
&=&\int_\gamma\d t\Biggl|{\partial p_\varphi\over\partial\varphi''}
  \Biggr|^{1/2}
  =\int_\gamma\d t\left|
  {\partial\varphi(t+T_\gamma)\over\partial p_\varphi(t)}
  \right|^{-1/2}~.
\end{eqnarray}


\newpage
\begin{table}
\caption{ \label{VARS}
Properties of the periodic orbits: periods $T$ (in unit of
$1/\osh(0)$) and traces of the reduced monodromy matrices $\Tr M$,
evaluated for $\loct=0$, 0.2, 0.3 and 0.4.  Here, `$n\gamma$' denote
the $n$-fold traversal of the primitive orbit $\gamma$.
For the isolated orbit A' and D, the monodromy matrix $M$ has two
unit-eigenvalues and the remaining four eigenvalues appear in pairs
$(e^{\alpha_{\rm a}}, e^{-\alpha_{\rm a}})$ and $(e^{\alpha_{\rm b}},
e^{-\alpha_{\rm b}})$.  These pairs are identical ($\alpha_{\rm
a}=\alpha_{\rm b}$) for orbit D, but they differ from each other for
orbit A'.  Traces of these pairs are given for orbit A'.
}
\bigskip

\def\bks{\!\!\!\!\!\!}
\def\bara{---}
\def\barb{---$\!$}
\def\barc{---~~~}

\def\cases#1#2{
  \Big\lbrace{
  \mbox{\hfill\small $#1$}\atop
  \mbox{\hfill\small $#2$}}
}

\begin{tabular}{
@{\quad}r@{\quad}
*{4}{|@{\quad}c@{\quad}r@{\quad}}}
$\lambda_{30}$
& \multicolumn{1}{r}{$0\!$}   &
& \multicolumn{1}{r}{$0.2\bks$} &
& \multicolumn{1}{r}{$0.3\bks$} &
& \multicolumn{1}{r}{$0.4\bks$} &
\\
\hline
orbit
& $T/\pi$ & $\Tr M\!\!$
& $T/\pi$ & $\Tr M$
& $T/\pi$ & $\Tr M$
& $T/\pi$ & $\Tr M$
\\
\hline
A
& 1     & $-2$
& 1.018 & $-1.778$
& 1.037 & $-1.362$
& 1.062 & $-0.758$
\\
2A
& 2     & 2
& 2.036 & 1.161
& 2.075 & $-0.145$
& 2.125 & $-1.426$
\\
3A
& 3     & $-2$
& 3.054 & $-0.286$
& 3.112 & 1.560
& 3.187 & 1.838
\\
E
& \bara & \barb
& \bara & \barc
& \bara & \barc
& 3.181 & 1.546
\\
F
& \bara & \barb
& \bara & \barc
& \bara & \barc
& 3.183 & 2.292
\\
B
& 2     & 2
& 1.999 & 1.816
& 1.998 & 1.017
& 1.995 & $-2.054$
\\
2B
& 4     & 2
& 3.999 & 1.300
& 3.997 & $-0.966$
& 3.989 & 2.221
\\
K
& \bara & \barb
& \bara & \barc
& \bara & \barc
& 3.989 & 1.572
\\
C
& 2     & 2
& 2.001 & 2.030
& 2.004 & 2.383
& 2.009 & 4.277
\\
D
& 2     & 2
& 2.072 & 1.845
& 2.177 & 3.498
& 2.367 & $-16.317$
\\
A'$\!$
& 1     & $\cases{-2}{2}\!$
& 1.024 & $\cases{-1.777}{1.992}\!$
& 1.050 & $\cases{-1.264}{1.960}\!$
& 1.086 & $\cases{-0.499}{1.869}\!$
\\
C'$\!$
& 2     & 2
& 2.003 & 2.046
& 2.008 & 2.333
& 2.016 & 3.484
\\
\end{tabular}
\end{table}

\begin{figure}[t]
\figcapt{SPSD}{
Results of semiclassical calculation of the oscillating part of the
level density for the SD oscillator model.
a) The contribution from orbits with $T=2\pi/\omega_z$,
b) the contribution from orbits with $T=2\pi/\omega_\perp$, and
c) the supershell structure caused by the interference between the
above two families of orbits.
}
\end{figure}

\begin{figure}[t]
\figcapt{OLD}{
Oscillating parts of the level density for the Hamiltonian
(\protect\ref{shamil}) with $\loct=0.2$, 0.3 and 0.4, calculated by
means of the Strutinsky method with the smoothing width
$\gamma=0.5\hbar\osh$.
The arrows indicate the minima associated with the even-$\Nsh$ closure.
}
\end{figure}

\begin{figure}[t]
\figcapt{SUPS}{
The oscillating part of the level density at $\loct=0.4$ obtained by
the Strutinsky method (broken line) is compared with a superposition
(solid line) of ~$\cos(ET_\gamma/\hbar)$ with the period $T_\gamma$
evaluated for the classical orbit A ($T\approx2\pi/\omega_\perp$) and
that for orbits B, C, C', and 2A ($T\approx2\pi/\osh$) (see \S5 for
the properties of these classical orbits).  Amplitudes and phases of
these cosine functions are determined so that the solid line best
agrees with the broken line, except that the energy dependence of the
amplitudes is assumed to fulfill the relation in
Eq.~(\protect\ref{scalerule}) determined by the scaling property of
the system under consideration (see \S6).
}
\end{figure}

\begin{figure}[t]
\figcapt{PMAPA}{
Poincar\'e maps in the section $(z,p_z)$ for the
Hamiltonian (\protect\ref{chamil}) with $p_\varphi=0$ and with
$\loct=0.2\sim0.4$, defined by $p_\rho=0$ and $\dot{p}_\rho<0$.
}
\end{figure}

\begin{figure}[t]
\figcapt{PMAPC}{
Poincar\'e maps in the section $(z,p_z)$ for the
Hamiltonian~(\protect\ref{chamil}) with $\loct=0.4$ and
with $p_\varphi/E=0.2$ and 0.4, defined by $p_\rho=0$ and
$\dot{p}_\rho<0$.}
\end{figure}

\begin{figure}[t]
\figcapt{PMAPB}{
Poincar\'e maps in the section $(\rho,p_\rho)$ for the
Hamiltonian~(\protect\ref{chamil}) with $p_\varphi=0$ and with
$\loct=0.2$ and 0.4, defined by $z=0$ and $p_z>0$.}
\end{figure}

\begin{figure}[t]
\figcapt{POC}{
Short periodic orbits for the Hamiltonian~(\protect\ref{chamil}) with
$\loct=0.4$.  Upper part: Planar orbits in the plane containing
the symmetric axis $z$.  Lower part: A circular orbit in the plane
perpendicular to the symmetry axis (A') and a three-dimensional orbit
(C').  Their projections on the $(x,y)$ plane and on the $(z,y)$ plane
are shown.}
\end{figure}

\begin{figure}[t]
\figcapt{POP}{
Short planar orbits for the Hamiltonian~(\protect\ref{chamil}) with
$p_\varphi=0$ and $\loct=0.3\sim0.4$.}
\end{figure}

\begin{figure}[t]
\figcapt{TRM}{
Traces of the reduced monodromy matrices $\Tr\tilde{M}$ for the
non-isolated periodic orbits shown in Fig.~\protect\ref{POC}
(see text for their definitions).
}
\end{figure}

\begin{figure}[t]
\figcapt{FTLA}{
Fourier transform $P_{\Delta s}^{\rm(qm)}(s)$ of the level density
$g(E)$ defined by Eq.~(\protect\ref{avqpower}), for $\loct=0.2$, 0.3
and 0.4.  Gaussian cut off is done with $E_{\rm max}\equiv\hbar/\Delta
s=15\hbar\osh(\loct)$.  Arrows indicate periods of the classical
periodic orbits (see Fig.~\protect\ref{POC}) and of their repetitions.
This figure is basically the same as Fig.~15 in our previous
report;\refmk{\protect\cite{MizNak})}  but, accuracy of the numerical
calculation is significantly improved so that the peak at
$s/\pi\approx1$ is now clearly seen.  This improvement greatly
facilitates the discussion on the classical-quantum correspondence
(see text).
}
\end{figure}

\begin{figure}[t]
\figcapt{FTO}{
Peaks heights in the Fourier transform defined by
Eq.~(\protect\ref{avqpower})
at the periods of classical periodic orbit A and of its multiple
traversals ($m$=1,3,5), plotted as functions of $\loct$.  Gaussian cut
off is done with $E_{\rm max}=12\hbar\osh(\loct)$.
Two arrows represent the period-tripling and period-5-upling
bifurcation points.
}
\end{figure}

\begin{figure}[t]
\figcapt{FTLZ}{
Fourier transforms of the level density $g(E,m)$ in the $m$=0 subspace
for $\loct$=0.2, 0.3 and 0.4.}
\end{figure}


\begin{references}
\bibitem{Dudek}
J. Dudek, T. T. Werner and Z. Szymanski,
Phys. Lett. {\bf B248} (1990), 235.

\bibitem{HolAb}
J. H\"oller and S. {\AA}berg,
Z. Phys. {\bf A336} (1990), 363.

\bibitem{Chasman}
R. R. Chasman,
Phys. Lett. {\bf B266} (1991), 243.

\bibitem{Li}
Xunjun Li, J. Dudek and P. Romain,
Phys. Lett. {\bf B271} (1991), 281.

\bibitem{Skalski}
J. Skalski,
Phys. Lett. {\bf B274} (1992), 1.

\bibitem{Bonche}
P. Bonche, S. J. Krieger, M. S. Weiss, J. Dobaczewski, H. Flocard
and P.-H. Heenen,
Phys. Rev. Lett. {\bf 66} (1991), 876.

\bibitem{SkalHe}
J. Skalski, P.-H. Heenen, P. Bonche, H. Flocard and J. Meyer,
Nucl. Phys. {\bf A551} (1993), 109.

\bibitem{NazDob}
W. Nazarewicz and J. Dobaczewski,
Phys. Rev. Lett. {\bf 68} (1992), 154.

\bibitem{Bengtsson}
T. Bengtsson, M. E. Faber, G. Leander, P. M\"oller, M. Ploszajczak,
I. Ragnarsson and S. {\AA}berg,
Physica Scripta {\bf 24} (1981), 200.

\bibitem{AriMat}
K. Arita and K. Matsuyanagi,
Prog. Theor. Phys. {\bf 89} (1993), 389.

\bibitem{MizNak}
S. Mizutori, T. Nakatsukasa, K. Arita, Y. R. Shimizu
and K. Matsuyanagi,
Nucl. Phys. {\bf A557} (1993), 125c.

\bibitem{Balian}
R. Balian and C. Bloch,
Ann. of Phys. {\bf 69} (1972), 76.

\bibitem{NisHanMot}
H. Nishioka, Klavs Hansen and B. R. Mottelson,
Phys. Rev. {\bf B42} (1990), 9377.

\bibitem{Nishioka}
H. Nishioka,
Z. Phys. {\bf D19} (1991), 19.

\bibitem{Brack}
M. Brack, J. Damgaard, A. S. Jensen, H. C. Pauli, V. M. Strutinsky
and C. Y. Wong,
Rev. Mod. Phys. {\bf 44} (1972), 320.

\bibitem{BohMot}
A. Bohr and B. R. Mottelson,
{\it Nuclear Structure} (Benjamin, 1975) Vol.~2, p.~585.

\bibitem{StrMag}
V. M. Strutinsky and A. G. Magner,
Sov. J. Part. Nucl. {\bf 7} (1976), 138.

\bibitem{StrMag}
V. M. Strutinsky, A. G. Magner, S. R. Ofengenden
and T. D\o ssing,
Z. Phys. {\bf A283} (1977), 269.

\bibitem{Frisk}
H. Frisk,
Nucl. Phys. {\bf A511} (1990), 309.

\bibitem{Malta}
C. P. Malta, M. A. M. de Aguiar and A. M. Ozorio de Almeida,
Phys. Rev. {\bf A47} (1993), 1625.

\bibitem{Gutztext}
M. C. Gutzwiller,
{\it Chaos in Quantum and Classical Mechanics},
(Springer Verlag, 1990).

\bibitem{GutzLH}
M. C. Gutzwiller,
{\it The Semi-classical Quantization of Chaotic Hamiltonian Systems},
in: Chaos and Quantum Physics,
Proc. Les Houches Summer School, Session LII (1989), 201.

\bibitem{Ozorio}
A. M. Ozorio de Almeida,
{\it Hamiltonian System, Chaos and Quantizaion},
(Cambridge University Press, 1988).

\bibitem{Arita}
K. Arita,
Prog. Theor. Phys. {\bf 90} (1993), 747.

\bibitem{BerTab}
M. V. Berry and  M. Tabor,
Proc. Roy. Soc. Lond. {\bf A349} (1976), 101.

\bibitem{Gutz}
M. C. Gutzwiller,
J. Math. Phys. {\bf 8} (1967), 1979;
{\bf 12} (1971), 343.

\bibitem{BarDav}
M. Baranger, K. T. R. Davies and J. H. Mahoney,
Ann. of Phys. {\bf 186} (1988), 95.

\bibitem{OzoHan}
A. M. Ozorio de Almeida and J. H. Hannay,
J. of Phys. {\bf A20} (1987), 5873.

\bibitem{Magner}
A. G. Magner,
Sov. J. Nucl. Phys. {\bf 28} (1978), 759.

\bibitem{SakKis}
H. Sakamoto and T. Kishimoto,
Nucl. Phys. {\bf A501} (1989), 205.

\bibitem{Wintgen}
D. Wintgen,
Phys. Rev. Lett. {\bf 58} (1987), 1589;
{\bf 61} (1988), 1803.

\bibitem{MagKolStr}
A. G. Magner, V. M. Kolomietz and V. M. Strutinsky,
Sov. J. Nucl. Phys. {\bf 28} (1978), 764.
\end{references}
\end{document}